# Spin-torque effects in thermally assisted magnetization reversal: Method of statistical moments

Y.P. Kalmykov,[1] W.T. Coffey,[2] S.V. Titov,[3] J.E. Wegrowe,[4] D. Byrne[5]

[1]*Laboratoire de Mathématiques et Physique, Université de Perpignan Via Domitia, F-66860, Perpignan, France*

[2]*Department of Electronic and Electrical Engineering, Trinity College, Dublin 2, Ireland*

[3]*Kotel'nikov Institute of Radio Engineering and Electronics of the Russian Academy of Sciences, Vvedenskii Square 1, Fryazino, Moscow Region, 141120, Russia*

[4]*Laboratoire des Solides Irradiés, Ecole Polytechnique, 91128 Palaiseau Cedex, France*

[5]*School of Physics, University College Dublin, Belfield, Dublin 4, Ireland*

**Abstract**

Thermal fluctuations of nanomagnets driven by spin-polarized currents are treated via the Landau-Lifshitz-Gilbert equation generalized to include both the random thermal noise field and the Slonczewski spin-transfer torque term. By averaging this stochastic (Langevin) equation over its realizations, the *explicit* infinite hierarchy of differential-recurrence relations for statistical moments (averaged spherical harmonics) is derived for *arbitrary* demagnetizing factors and magnetocrystalline anisotropy for the generic nanopillar model of a spin-torque device comprising two ferromagnetic strata representing the free and fixed layers and a nonmagnetic conducting spacer all sandwiched between two ohmic contacts. The influence of thermal fluctuations and spin-transfer torques on relevant switching characteristics, such as the stationary magnetization, the magnetization reversal time, etc., is calculated by solving the hierarchy for wide ranges of temperature, damping, external magnetic field, and spin-polarized current indicating new spin-torque effects in the thermally assisted magnetization reversal comprising several orders of magnitude. In particular, a pronounced dependence of the switching characteristics on the directions of the external magnetic field and the spin polarization exists.

## I. INTRODUCTION

One of the most significant developments in the thermally assisted magnetization reversal in nanomagnets since the seminal work of Néel[1] and Brown[2,3] on the reversal time of the magnetization of single-domain ferromagnetic nanoparticles due to thermal fluctuations has been the discovery of the spin-transfer torque (STT) effect.[4-6] The phenomenon exists because an electric current with spin polarization in a small (nanoscale) ferromagnet may transfer spin angular momentum between the current and the magnetization **M** giving rise to the macroscopic spin-torque on **M**[4-7] so that the latter *may be altered by spin polarized currents*.[8] Such effects underpin the relatively new subject of spintronics,[9] where the carrier of information is the spin state of a ferromagnetic material. Typical practical applications include very-high-speed current-induced magnetization switching by (a) reversing the orientation of magnetic bits in high density memory structures as opposed to the more conventional Oersted field switching[7,10,11] and (b) using spin polarized currents both to generate and manipulate steady state microwave oscillations with a frequency proportional to the spin-polarization current[12,13] via the steady state magnetization precession. Essentially both objectives (a) and (b) can be achieved because, depending on its sign, the spin current either enhances or diminishes the effective damping representing the microscopic degrees of freedom of a ferromagnetic film[8] (cf. Eq. (1) below). The meaning of this,[8] considering a bistable potential, is that during a precessional period in a well, the average rate of change of energy $\dot{E}$ may be either negative, positive, or indeed zero. If $\dot{E} < 0$ the magnetization is forced to relax into its energy minimum in the well. On the other hand, if $\dot{E}$ is equal to zero, we have stable precession at constant energy as if the Gilbert damping were absent. Finally if $\dot{E} > 0$, resulting in very large precessional orbits at energies in the vicinity of the saddle energy, the magnetization is ultimately forced to switch to a new stable position in the other well of the potential (see Fig. 2 below).

Regarding objective (b) above, a simple treatment of the onset of stable precessional states at zero temperature has here given in Ref. 6. There, since the damping torque is small and roughly balances the STT while ignoring thermal noise (represented by a stochastic magnetic field), perturbation theory is used to investigate the onset of precessional states. This is accomplished by studying those relatively low-energy phase-space trajectories on which the magnetization is close to its stable directions. Thus the equation of motion of the magnetization (cf. Eq. (1) below) may be linearized, as is usual in the theory of small oscillations,[14] about a stable direction. This leads to situations, where precessional motion exponentially decays for currents less than the critical value for the onset of precessional states [not to be confused with the switching current in objective (a)[6]] and exponentially grows for currents exceeding that value. The phenomenon constitutes a parametric excitation because the STT is a function of the orientation of **M**, representing a time varying modification of a system parameter and thus may exhibit instability unlike conventional resonance.[15] Indeed the overall behavior is more or less analogous to that of a triode vacuum tube oscillator[16] whereby a coil connected to the grid circuit is coupled via mutual inductance to a lightly damped oscillatory circuit in the anode circuit. Then while the triode is in operation, the resulting feedback effect is either to decrease or increase the effective resistance of the oscillatory anode circuit according to the sense in which the coil in the grid is connected. If the damping is decreased and the mutual coupling is sufficient, the former may be reduced to zero. Thus an oscillation once started will persist and will grow until limited by the characteristics of the tube. A discussion of this limiting behavior in the spin-torque case is given in Ref. 6.

Now regarding objective (a) because[6] the STT represents a parametric excitation with $\dot{E}$ >0 then bifurcation phenomena due to parametric amplification at energies in the vicinity of the barrier energy may manifest themselves whereby a qualitatively different solution for a nonlinear system may appear following the variation of some parameter. In our context, this behavior represents crossing of the potential barrier causing the magnetization to evolve into more highly damped precessional states exhibiting ringing oscillations which decay rapidly (see Fig. 2). Thus the magnetization is driven into its new stationary state, where precession is prevented due to the sign of the STT which now *enhances* rather than *reduces* the damping. However, unlike the value of the critical current characterizing the *onset* of precessional states, originating in the small oscillations about a stationary orientation of the magnetization, no closed form[6] for the switching current (at which the direction of precession reverses) can be derived from simple perturbation theory. This is so because unlike the *small* oscillations, the switching has its origin in the *large* oscillations about the direction of precession characterizing the motion near the saddle (barrier) energy between two stable states. Here the precession is almost metastable and so may easily be reversed in direction following a small change in the energy (see Fig. 2). An essentially similar argument was used by Kramers.[17] He utilized the concept of large oscillations of Brownian particles in a well (at energies near the separatrix energy between the bounded motion in the well and the unbounded motion outside it) in discussing noise-induced escape for very weak coupling to a thermal bath as explained in Ref. 18. In the STT application, any results that are available utilize Melnikov's method[13, 19] for weakly perturbed Hamiltonian systems that are periodic in time, where the unperturbed trajectories in phase space may be derived from the energyscape. One of the major benefits of his method is that it establishes[6] a clear distinction between the critical currents for the onset of precessional states and those for switching.

All of the foregoing discussion pertains to zero temperature, where, for example,[6] the precessional states are characterized in the single macrospin approximation by a frequency that is a function of the current density, the external magnetic field, the anisotropy, the damping parameter, etc. However, the thermal fluctuations cannot be ignored due to the *nanometric* size of STT devices, e.g., leading to mainly noise-induced switching at currents far less than the critical switching current without noise[20] a phenomenon corroborated by experiments (e.g., Ref. 21) which demonstrate that STT near room temperature alters thermally activated switching processes. At finite temperatures, randomness due to the thermal motion of the surroundings is introduced into the magnetization trajectories, counteracting the damping and giving rise to fluctuations as compared to the zero temperature limit. Furthermore, large fluctuations can cause transitions between metastable states[6] of the magnetization at currents less than the zero temperature current essentially in the manner envisaged by Kramers.[17,18] Here we study the effects of thermal fluctuations in the presence of STT using an adaptation of the Langevin equation for the evolution of a single macrospin proposed by Brown.[2,3] He showed how the noise-induced magnetization relaxation, i.e., reversal of the direction of precession by crossing over a potential barrier between two equilibrium states could be set firmly within the context of the theory of stochastic processes. However, it should be recalled throughout that unlike in the original work of Brown and Néel[1-3] STT devices, due to the injection of the spin-polarized current, invariably represent an *open* system in an *out-of-equilibrium steady state.* Such behavior is in marked contrast to the conventional steady state of nanostructures characterized by the Boltzmann equilibrium distribution that arises when the STTs are omitted. Now, the effect of thermal fluctuations, using a modification of the customary Néel-Brown model,[1-3] treated here represent a very significant feature of their operation. Fluctuations, for example, lead to mainly noise-induced switching at currents far less than the critical current in the absence of noise as well as introducing randomness into the precessional orbits which now exhibit energy-controlled diffusion. Thus the effect of the noise is generally to reduce the current-induced switching time.[20] To facilitate our discussion we first describe the archetypal schematic model of the STT effect.

## II. MODEL

The archetypal model (Fig. 1) of a STT device is a nanostructure consisting of two magnetic strata labeled the *free* and *fixed* layers, respectively, and a nonmagnetic conducting spacer all sandwiched on a pillar between two ohmic contacts.[6,13] The fixed and free layers differ significantly because the fixed layer is pinned[12] along its orientation much more strongly than the free one usually because the former is of a harder magnetic material so that the latter is much easier to manipulate in a magnetic sense. When an electric current is passed through the fixed layer, it becomes polarized. The polarized spin current then encounters the free layer and induces a spin torque altering its magnetization so permitting[8] a variety of dynamical regimes. This phenomenon can lead, in particular, to two distinct magnetization dynamics regimes which have been extensively verified experimentally,[20] viz., steady-state precession and STT-induced reversal of the direction of precession governed by the transition rate between precessional states. Consequently, one can introduce two distinct time scales associated with the magnetization vector **M**, namely, a *slow* one, corresponding to reversal of the magnetization over a potential barrier and a *fast* one, characterized by the precession frequency of the bounded motion in a potential well for constant energy.

Now in the model under consideration both ferromagnetic layers are assumed to be uniformly magnetized (for small ferromagnets the STT may lead to a rotation of the magnetization as a whole, rather than to an excitation of spin waves; even though the single-domain or coherent rotation approximation cannot explain all observations of the magnetization dynamics in spin-torque systems, many qualitative features needed to explain experimental data are satisfactorily reproduced[6,13]). Thus in the presence of thermal fluctuations, the current-induced magnetization dynamics of **M** in the free layer is governed by the Slonczewski equation (i.e., the Landau-Lifshitz-Gilbert equation[3] including the spin-torque)[13] augmented by a random magnetic field **h** with Gaussian white noise properties and so becoming a Langevin equation[4,6,13,22]

$$\dot{\mathbf{u}} = -\gamma \left[ \mathbf{u} \times \left( \mathbf{H}_{\text{ef}} + \dot{\mathbf{u}}\big|_{\text{ST}} + \mathbf{h} \right) \right] + \alpha \left[ \mathbf{u} \times \dot{\mathbf{u}} \right]. \qquad (1)$$

The Gaussian field **h** has the usual white noise properties

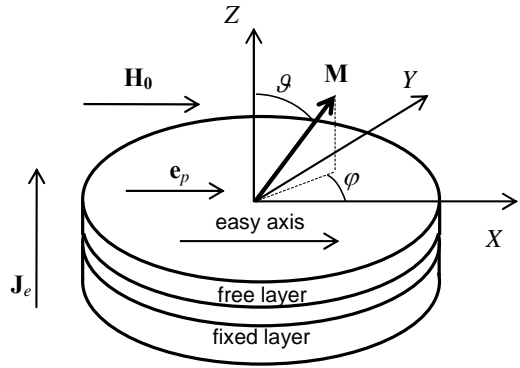

FIG. 1. Geometry of the problem. A STT device consists of two ferromagnetic strata labelled the *free* and *fixed* layers, respectively, and a normal conducting spacer all sandwiched on a pillar between two ohmic contacts.[13] The fixed layer has a fixed magnetization along the direction $\mathbf{e}_P$. $\mathbf{J}_e$ is the spin-polarized current density, **M** is the magnetization of the free layer, and $\mathbf{H}_0$ is the applied magnetic field.

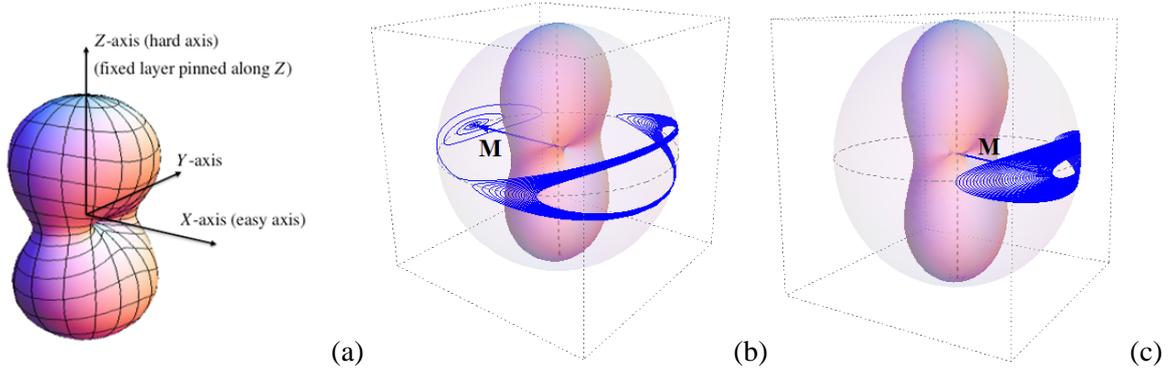

(a) (b) (c)

FIG. 2. (a) Biaxial anisotropy potential $V(\vartheta,\varphi)$, Eq. (3). (b) Current-induced trajectory of the magnetization escape. Solid line: numerical solution of the deterministic Eq. (1), i.e., omitting the random field **h**, for a strong spin-polarized current, damping $\alpha = 0.01$, and typical values of other model parameters (see Sec. V). Reversal of the magnetization from one metastable state to another typically occurs after many precessions about the *X* axis, over saddle points. Thus having traversed the potential barrier, the magnetization decays to a new stable direction of precession so that in this reverse direction the current accelerates the decay by increasing the effective damping, as obvious from Eq. (1). (c) The same as in Fig. 2(b) for a weak spin-polarized current and the same values of other model parameters. Here, only the damped precessions of the magnetization about the stable direction exist.

$$\overline{h_i(t)} = 0, \quad \overline{h_i(t_1)h_j(t_2)} = \frac{2kT\alpha}{v\gamma\mu_0 M_S}\delta_{i,j}\delta(t_1-t_2). \qquad (2)$$

where the indices $i,j = 1,2,3$ in Kronecker's delta $\delta_{i,j}$ and $h_i$ correspond to the Cartesian axes *X,Y,Z* of the laboratory coordinate system *OXYZ*, and $\delta(t)$ is the Dirac-delta function. The overbar means the statistical average over an ensemble of moments which all have at time *t* the *same* sharp value of the magnetization **M**, the sharp values subsequently being regarded as random variables. In Eq. (1), $\mathbf{u} = M_S^{-1}\mathbf{M}$ is a unit vector in the direction of **M**, $M_S$ is the saturation magnetization, $\gamma$ is the gyromagnetic-type constant, $\alpha$ is a damping parameter representing the effect of all the microscopic degrees of freedom, $\dot{\mathbf{u}}|_{ST}$ is the STT, $\mathbf{H}_{ef} = -(\mu_0 M_S)^{-1}\partial V/\partial \mathbf{u}$ is the effective magnetic field comprising the anisotropy and external fields, and the operator $\partial/\partial \mathbf{u}$ indicates the gradient on the surface of the unit sphere. Here *V*, constituting a conservative potential, is the normalized free energy per unit volume of the free layer which we write in the standard form of superimposed easy-plane and in-plane easy-axis anisotropies plus the Zeeman term, viz.,[13,22]

$$V(\vartheta,\varphi) = \mu_0 M_S^2 D_\parallel \big[\delta\cos^2\vartheta - \sin^2\vartheta\cos^2\varphi \\ -2h(\gamma_1\cos\varphi\sin\vartheta + \gamma_2\sin\varphi\sin\vartheta + \gamma_3\cos\vartheta)\big] \qquad (3)$$

as shown in Fig. 2a. Thus the potential creates an energyscape with two minima and two saddle points and forces the magnetization to align in a given direction in one or other of the energy minima in the equatorial or *XY* plane. Here the *Z* axis is obviously taken as the hard axis while the *X* axis is the easy one; $\vartheta$ and $\varphi$ are angular coordinates describing the orientation of the moment **u** in the spherical polar coordinate system, $\gamma_1 = \cos\varphi_\xi \sin\vartheta_\xi$, $\gamma_2 = \sin\varphi_\xi \sin\vartheta_\xi$, and $\gamma_3 = \cos\vartheta_\xi$ are the direction cosines of the applied field $\mathbf{H}_0$, $h = H_0/(2M_S D_\parallel)$ is the external field parameter, and $\delta = D_\perp/D_\parallel$ is the biaxiality parameter, where $D_\parallel$ and $D_\perp$ account for both demagnetizing and magnetocrystalline anisotropy effects.[13] The spin-transfer torque term $\dot{\mathbf{u}}|_{ST}$ is defined as

$$\dot{\mathbf{u}}|_{ST} = -\frac{1}{\mu_0 M_S}\left[\mathbf{u}\times\frac{\partial\Phi}{\partial\mathbf{u}}\right],$$

where $\Phi$ is the non-conservative potential due to the spin-polarized current given by [4]

$$\Phi(\mathbf{u}) = \frac{\mu_0 M_S^2 b_P J_e}{c_P J_p}\ln(1 + c_P \mathbf{u}\cdot\mathbf{e}_P). \tag{4}$$

In Eq. (4) the unit vector $\mathbf{e}_P$ identifies the magnetization direction in the fixed layer, cf. Fig.1, $J_e$ is the current density, taken as positive when the electrons flow from the free into the fixed layer, while $J_p = \mu_0 M_S^2 |e| d/\hbar$ ($e$ is the electronic charge, $\hbar$ is Planck's reduced constant, and $d$ is the thickness of the free layer). The coefficients $b_P$ and $c_P$ are model dependent and are determined by the spin-polarization factor $P(0 < P < 1)$ [4]

$$b_P = \frac{4P^{3/2}}{3(1+P)^3 - 16P^{3/2}},$$

$$c_P = \frac{(1+P)^3}{3(1+P)^3 - 16P^{3/2}}$$

where $0 < b_P < 1/2$ and $1/3 < c_P < 1$ as $P$ increases from 0 to 1. The typical value of $J_p$ for a 3 nm- thick layer of cobalt, where $M_S \approx 1.4\cdot 10^6$ A m$^{-1}$, is $J_p \approx 1.1\cdot 10^9$ A/cm$^2$ (cf. Ref. 13, p. 237).

In tandem with Eq. (1), we have[13] (see Sec. 1.17 of Ref. 23) the Fokker-Planck equation (FPE) for the probability density function $W(\vartheta,\varphi,t)$ of orientations of **u** on the unit sphere, viz.,

$$2\tau_N \frac{\partial W}{\partial t} = \frac{1}{\sin\vartheta} \frac{\partial}{\partial \vartheta} \left\{ \sin\vartheta \left[ \frac{\partial W}{\partial \vartheta} + \frac{v}{kT} W \left( \frac{\partial (V+\alpha^{-1}\Phi)}{\partial \vartheta} + \frac{1}{\sin\vartheta} \frac{\partial (\alpha^{-1}V-\Phi)}{\partial \varphi} \right) \right] \right\}$$
$$+ \frac{1}{\sin^2\vartheta} \frac{\partial}{\partial \varphi} \left[ \frac{\partial W}{\partial \varphi} + \frac{v}{kT} W \left( \frac{\partial (V+\alpha^{-1}\Phi)}{\partial \varphi} - \sin\vartheta \frac{\partial (\alpha^{-1}V-\Phi)}{\partial \vartheta} \right) \right]. \quad (5)$$

Here $\tau_N = v\mu_0 M_S(\alpha + \alpha^{-1})/(2\gamma kT)$ is the free diffusion time of the magnetic moment, $kT$ is the thermal energy, and $v$ is the free layer volume.

By way of background to the Langevin equation method, during the last decade, various analytical and numerical approaches to the calculation of the measurable parameters from the Langevin and Fokker-Planck equations (1) and (5) have been extensively developed, e.g., generalizations (e.g., Refs. 24-26) of the Kramers escape rate theory[17] and stochastic dynamics simulations (e.g., Refs. 24, 27-29). For example, the pronounced time separation between *fast* precessional and *slow* energy changes in lightly damped ($\alpha \ll 1$) closed phase space trajectories (called Stoner-Wohlfarth orbits) at energies near the barrier energy has been exploited in Ref. 12 to formulate a FPE for the energy distribution essentially similar to that of Kramers[17] for point particles. Regarding the magnetization reversal time it will become apparent that the effect of the spin-polarized current may be of several orders of magnitude in the very-low-damping limit ($\alpha \ll 1$, the only relevant case) since the stationary distribution of orientations of **M** is no longer the Boltzmann one as it now depends both on the spin polarized current $J_e$ and on $\alpha$. The dependence is all the more obvious when one considers, just as in Ref. 12, the stationary solution of the Fokker-Planck equation for the axially symmetric $V$ and $\Phi$ which arises for uniaxial anisotropy with the easy axis, the magnetization direction in the fixed layer, and the external field taken as collinear. In this special case alone the magnetization dynamics are determined by a simple generalized potential yielding the stationary distribution in exact closed form as well as an approximate expression for the reversal time. The *effective* potential $V + \alpha^{-1}\Phi$ comprises that of the conservative external and anisotropy fields as well as the *nonconservative* one due to the spin-polarized current. In general, the existence of a nonconservative effective potential[12,13,29] allows one to define a current-dependent potential barrier between stationary self-oscillatory states (limit cycles) of the magnetization and to estimate transition rates between these states. Now for axial symmetry the approximate solution procedure for the smallest nonvanishing eigenvalue $\lambda_1$ of the Fokker-Planck equation for axially symmetric potentials and for high barrier heights given by Brown[2,3] may be used. Here the asymptotic calculation of $\lambda_1$, thus the magnetization reversal time $\tau = 1/\lambda_1$, may be effected[2,3] via the purely mathematical method of approximate minimization (stemming from the calculus of variations) for $\lambda_1$ of the axially symmetric Fokker-Planck equation when converted to a Sturm-Liouville equation.

However, for *nonaxially symmetric* problems, as depicted in Fig. 2, no such simple asymptotic solution exists because it is impossible, by inspection of the nonaxially symmetric Fokker-Planck Eq. (5) when the STT is included, to derive a simple analytic equation for an effective potential. [Such a potential can be calculated only in numerically via the time-independent distribution function; see Eq. (27) below.] Nevertheless, it is still possible to calculate $\lambda_1$ in numerical fashion, once again yielding the magnetization reversal time. Moreover, it is also possible to calculate the time-independent in-plane component of **M**, i.e., in the *X* or easy-axis direction $\langle u_X \rangle_0$ as well as the corresponding magnetic susceptibility $\propto \langle u_X^2 \rangle_0 - \langle u_X \rangle_0^2$, where the angular brackets $\langle \ \rangle_0$ mean stationary statistical averaging. The merit of such calculations is that *inter alia* they allow one to accurately assess approximate low-damping solutions for the reversal time based on energy-controlled diffusion.[13,29] These solutions all more or less rest on (noting the separation of time scales referred to above) treating both the effects of the stochastic torques due to the heat bath and the spin torque as perturbations of the precessional dynamics of **M** in the wells of the anisotropy-Zeeman energy potential. The corresponding closed phase-space trajectories are known[8,20] as Stoner-Wohlfarth orbits and steady precession along such an orbit of constant energy, belonging to a sphere of radius equal to the saturation magnetization, occurs if the spin-torque cancels out the dissipative torque [cf. Eq. (1)]. The origin of the orbits of course arises from the two well structure of the anisotropy potential. One should at this juncture mention the treatment of reversal of the magnetisation by Apalkov and Visscher.[25] Here, for example, the time separation between the fast precessional and slow energy change in a lightly damped Stoner-Wohlfarth orbit at energies near to the barrier energy is used to formulate a Fokker-Planck equation for the probability distribution of the energy near the barrier. This method is again essentially similar to the approach used by Kramers[17,18] in the problem of the very-low-damping noise-activated escape rate from a potential well. Moreover, the derivation of the Fokker-Planck equation in energy-phase variables for point particles with separable and additive Hamiltonians has been extensively discussed by Stratonovich[30] and Risken.[31] The corresponding magnetic problem (where the Hamiltonian in the absence of spin torque is non-separable) in energy-precession variables has been discussed by Dunn *et al.*[8] in relation to their Eq. (1.22), which, on assuming rapid equilibration of the precession variable, leads to an energy diffusion equation, their Eq. (1.23).

In this paper, we shall present results for the stationary magnetization $\langle u_X \rangle_0$ and static susceptibility $\propto \langle u_X^2 \rangle_0 - \langle u_X \rangle_0^2$ in the easy-axis direction and for the reversal time of the magnetization. These results are obtained by extending the general method (based on the relevant Langevin equation) of constructing recurrence relations for the evolution of statistical moments for arbitrary nonaxially symmetric free energy given in Refs. 23 and 32 to include the STT term and then specializing them to Eq. (3). The recurrence relations can then be solved by matrix continued fraction methods just as with the zero STT term.[23] The answers will then constitute benchmarks for approximate solutions obtained by other methods. Indeed the procedure is entirely analogous to that involving the numerical solutions[3] used to test asymptotic solutions, based on the Kramers escape rate theory,[3,17] for the reversal time in the Néel-Brown model when the STT is absent. Notice that if the STT is included its effect may also be described via a modification of the energy barrier in the Néel-Brown model (for detailed discussions see Refs. 13, 24-26, and 33).

### III. DIFFERENTIAL-RECURRENCE RELATION FOR THE STATISTICAL MOMENTS

Our method[23,32] is based on first averaging the appropriate Langevin equation including the spin-torque term (regarded as a Stratonovitch stochastic differential equation) for the magnetization evolution, therefore written in terms of the spherical harmonics, over its realizations in the representation space of polar angles in an infinitesimally small time starting from a set of sharp angles, which subsequently are also regarded as random variables. The result is then averaged over the distribution of these angles ultimately yielding the desired recurrence relation for the observables which are the statistical moments of the system. Here the relevant Langevin equation is the Landau-Lifshitz-Gilbert equation for the evolution of **M** in the free layer as modified by Slonczewski[4,13] to include the STT, Eq. (1). As shown in Appendix A, Eq. (1) can be written in an equivalent Landau-Lifshitz form, where the precessional and alignment terms are now clearly delineated, viz.

$$\dot{\mathbf{u}} = \frac{v}{2kT\alpha\tau_N} \left[ \mathbf{u} \times \left[ \frac{\partial}{\partial \mathbf{u}}(V - \alpha\Phi) - \mu_0 M_S \mathbf{h} \right] \right] \\ + \frac{v}{2kT\tau_N} \left[ \mathbf{u} \times \left[ \mathbf{u} \times \left( \frac{\partial}{\partial \mathbf{u}}(V + \alpha^{-1}\Phi) - \mu_0 M_S \mathbf{h} \right) \right] \right]. \quad (6)$$

In the spherical polar coordinate basis shown in Fig. 1, the vector Langevin equation (6) represents two coupled nonlinear stochastic differential equations for the angles $\vartheta$ and $\varphi$, viz.,[23]

$$\dot{\vartheta}(t) = \frac{v\mu_0 M_S}{2kT\tau_N}\left[h_\vartheta(t)+\alpha^{-1}h_\varphi(t)\right] - \frac{v}{2kT\tau_N}\left\{\frac{\partial}{\partial\vartheta}\left(V[\vartheta(t),\varphi(t),t]+\alpha^{-1}\Phi[\vartheta(t),\varphi(t),t]\right)\right.$$
$$\left.+\frac{\alpha^{-1}}{\sin\vartheta(t)}\frac{\partial}{\partial\varphi}\left(V[\vartheta(t),\varphi(t),t]-\alpha\Phi[\vartheta(t),\varphi(t),t]\right)\right\},\tag{7}$$

$$\dot{\varphi}(t) = \frac{v\mu_0 M_S}{2kT\tau_N}\frac{h_\varphi(t)-\alpha^{-1}h_\vartheta(t)}{\sin\vartheta(t)} - \frac{v}{2kT\tau_N}\left\{\frac{1}{\sin^2\vartheta(t)}\frac{\partial}{\partial\varphi}\left(V[\vartheta(t),\varphi(t),t]+\alpha^{-1}\Phi[\vartheta(t),\varphi(t),t]\right)\right.$$
$$\left.-\frac{\alpha^{-1}}{\sin\vartheta(t)}\frac{\partial}{\partial\vartheta}\left(V[\vartheta(t),\varphi(t),t]-\alpha\Phi[\vartheta(t),\varphi(t),t]\right)\right\},\tag{8}$$

where the components $h_\vartheta(t)$ and $h_\varphi(t)$ of the Gaussian random field $\mathbf{h}(t)$ in the spherical basis are expressed in terms of the components $h_X(t)$, $h_Y(t)$, $h_Z(t)$ in the Cartesian basis as[23]

$$h_\vartheta(t) = h_X(t)\cos\vartheta(t)\cos\varphi(t) + h_Y(t)\cos\vartheta(t)\sin\varphi(t) - h_Z(t)\sin\vartheta(t),$$

$$h_\varphi(t) = -h_X(t)\sin\varphi(t) + h_Y(t)\cos\varphi(t).$$

Equations (7) and (8) then yield the desired Langevin equation for the evolution of the spherical harmonics $Y_{l,m}(\vartheta,\varphi)$[34] comprising the orthonormal basis set from which the observables are ultimately obtained, viz.,

$$\frac{dY_{l,m}}{dt} = \frac{\partial Y_{l,m}}{\partial\vartheta}\frac{d\vartheta}{dt} + \frac{\partial Y_{l,m}}{\partial\varphi}\frac{d\varphi}{dt}$$
$$= \frac{v\mu_0 M_S}{2kT\tau_N}\left\{\left[h_\vartheta(t)+\alpha^{-1}h_\varphi(t)\right]\frac{\partial Y_{l,m}}{\partial\vartheta} + \frac{h_\varphi(t)-\alpha^{-1}h_\vartheta(t)}{\sin\vartheta}\frac{\partial Y_{l,m}}{\partial\varphi}\right\}$$
$$-\frac{v}{2kT\tau_N}\left\{\left[\frac{\partial}{\partial\vartheta}\left(V+\alpha^{-1}\Phi\right)+\frac{1}{\sin\vartheta}\frac{\partial}{\partial\varphi}\left(\alpha^{-1}V-\Phi\right)\right]\frac{\partial Y_{l,m}}{\partial\vartheta}\right.$$
$$\left.+\left[\frac{1}{\sin^2\vartheta}\frac{\partial}{\partial\varphi}\left(V+\alpha^{-1}\Phi\right)-\frac{1}{\sin\vartheta}\frac{\partial}{\partial\vartheta}\left(\alpha^{-1}V-\Phi\right)\right]\frac{\partial Y_{l,m}}{\partial\varphi}\right\},\tag{9}$$

where $Y_{l,m}(\vartheta,\varphi)$ are defined by[34]

$$Y_{l,m}(\vartheta,\varphi) = \sqrt{\frac{(2l+1)(l-m)!}{4\pi(l+m)!}}e^{im\varphi}P_l^m(\cos\vartheta),$$

$$Y_{l,-m} = (-1)^m Y_{l,m}^*,$$

$P_l^m(x)$ are the associated Legendre functions,[34] and the asterisk denotes the complex conjugate.

By averaging this Langevin equation (9) as explained in Sec. 9.2 of Ref. 23 and summarized in Appendix B, we have the evolution equation of the statistical moments $\langle Y_{l,m}\rangle(t)$ (expected values of the spherical harmonics $Y_{l,m}$[34]) for *arbitrary* anisotropy rendered as the differential-recurrence relation viz.,

$$\tau_N \frac{d}{dt}\langle Y_{l,m}\rangle(t) = \sum_{l',m'} e_{l,m,l',m'} \langle Y_{l',m'}\rangle(t). \tag{10}$$

In Eq. (10) angular brackets mean statistical averaging and $e_{l,m,l',m'}$ are expressed via the Clebsch–Gordan coefficients $C^{L,M}_{l,m,l',m'}$ (Ref. 34) as

$$\begin{aligned}
e_{l,m,l',m\pm s} = & -\frac{l(l+1)}{2}\delta_{l,l'}\delta_{s,0} + (-1)^m \frac{1}{4}\sqrt{\frac{(2l+1)(2l'+1)}{\pi}} \\
& \times \sum_{r=s}^{\infty}\left\{\left(A_{r,\pm s} + \frac{B_{r,\pm s}}{\alpha}\right)\frac{\left[l'(l'+1)-r(r+1)-l(l+1)\right]}{2\sqrt{2r+1}}C^{r,0}_{l,0,l',0}C^{r,\mp s}_{l,m,l',-m\mp s} \right. \\
& \qquad -\frac{i}{\alpha}\left(A_{r,\pm s}-\alpha B_{r,\pm s}\right)\sqrt{\frac{(2r+1)(r-s)!}{(r+s)!}} \\
& \left. \times \sum_{\substack{L=s-\varepsilon_{r,s},\\ \Delta L=2}}^{r-1}\sqrt{\frac{(L+s)!}{(L-s)!}}C^{L,0}_{l,0,l',0}\left[mC^{L,\mp s}_{l,m,l',-m\mp s}\pm s\sqrt{\frac{(l\mp m)(l\pm m+1)}{(L+s)(L-s+1)}}C^{L,\mp s\pm 1}_{l,m\pm 1,l',-m\mp s}\right]\right\},
\end{aligned} \tag{11}$$

where $s \geq 0$ and $A_{r,s}$ and $B_{r,s}$ are, respectively, the coefficients of the Fourier series expansions in terms of spherical harmonics of the (conservative) free energy density $V$ and the nonconservative potential $\Phi$, viz.,

$$\frac{vV}{kT} = \sum_{r,s} A_{r,s} Y_{r,s}, \tag{12}$$

$$\frac{v\Phi}{kT} = \sum_{r,s} B_{r,s} Y_{r,s}. \tag{13}$$

Equation (10) represents the evolution of a typical entry in a set of differential-recurrence relations with $e_{l,m,l',m'}$ given by Eq. (11). Only the Fourier expansions of both potentials in terms of $A_{r,s}$ and $B_{r,s}$ are needed. In the *stationary* state, when the statistical moments are independent of time, Eq. (10) becomes

$$\sum_{l',s} e_{l,m,l',m'}\langle Y_{l',m'}\rangle_0 = 0, \tag{14}$$

The same result may be obtained, albeit with more labor and in a less transparent manner, from the Fokker-Planck equation, Eq. (5), by seeking the surface density of magnetic moment orientations on the unit sphere as[3,23,32]

$$W(\vartheta,\varphi,t) = \sum_{l=0}^{\infty}\sum_{m=-l}^{l} Y^*_{l,m}(\vartheta,\varphi)\langle Y_{l,m}\rangle(t), \tag{15}$$

where

$$\langle Y_{l,m}\rangle(t) = \int_0^{2\pi}\int_0^{\pi} Y_{l,m}(\vartheta,\varphi)W(\vartheta,\varphi,t)\sin\vartheta \, d\vartheta \, d\varphi \tag{16}$$

by the orthogonality property of the spherical harmonics, viz.,

$$\int_0^{2\pi}\int_0^{\pi} Y^*_{l_1,m_1}(\vartheta,\varphi)Y_{l_2,m_2}(\vartheta,\varphi)\sin\vartheta \, d\vartheta \, d\varphi = \delta_{l_1,l_2}\delta_{m_1,m_2}.$$

Equations (10) and (14) have been obtained under the assumption that the damping parameter $\alpha$ is independent of **M**. However, the results may be also generalized to magnetization-dependent damping $\alpha(\mathbf{M})$.[8,20]

Both Eq. (10) and Eq. (14) are valid for an *arbitrary* free energy. Here we specialize them to the particular free energy given by Eq. (3). In terms of spherical harmonics, Eq. (3) can be written as

$$\frac{vV}{kT} = -\frac{2\sigma}{3}\sqrt{\pi} + \sum_{r=1}^{2}\sum_{s=-r}^{r} A_{r,s} Y_{r,s}(\vartheta,\varphi), \tag{17}$$

where the nonzero expansion coefficients $A_{r,s}$ are given by

$$A_{1,0} = -4\sigma h \gamma_3 \sqrt{\frac{\pi}{3}},$$

$$A_{1,\pm 1} = \pm \sigma h (\gamma_1 \mp i\gamma_2)\sqrt{\frac{8\pi}{3}},$$

$$A_{20} = \sigma(1+2\delta)\sqrt{\frac{4\pi}{45}},$$

$$A_{2,\pm 2} = -\sigma\sqrt{\frac{2\pi}{15}},$$

and $\sigma = v\mu_0 M_S^2 D_\parallel/(kT)$ is an anisotropy (or inverse temperature) parameter. Now, the potential $\Phi$, Eq. (4), is in spherical harmonic notation

$$\frac{v\Phi}{kT} = J\frac{b_P}{c_P}\ln\left(1 + \frac{4\pi c_P}{3}\sum_{m=-1}^{1} Y_{1,m}(\vartheta,\varphi)Y_{1,m}^*(\vartheta_P,\varphi_P)\right), \tag{18}$$

where $\vartheta_P$ and $\varphi_P$ are the spherical polar coordinates of the unit vector $\mathbf{e}_p$ which is the magnetization direction of the fixed layer, and $J = v\mu_0 M_S^2 J_e/(kTJ_p)$ is the dimensionless spin-polarized current parameter. Next retaining only the two leading terms in the Taylor series expansion

$$\ln(1+x) = x - x^2/2 + x^3/3 - \ldots,$$

which converges fairly well for typical values of the model parameters, we will then have

$$\frac{v\Phi_{ap}}{kT} = \sum_{r=0}^{2}\sum_{s=-r}^{r} B_{r,s} Y_{r,s}(\vartheta,\varphi), \tag{19}$$

where the expansion coefficients $B_{r,s}$ are defined as

$$B_{0,0} = -\frac{4\pi^{3/2} b_P c_P J}{9}\left[Y_{1,0}^{*2}(\vartheta_P,\varphi_P) - 2Y_{1,1}^*(\vartheta_P,\varphi_P)Y_{1,-1}^*(\vartheta_P,\varphi_P)\right],$$

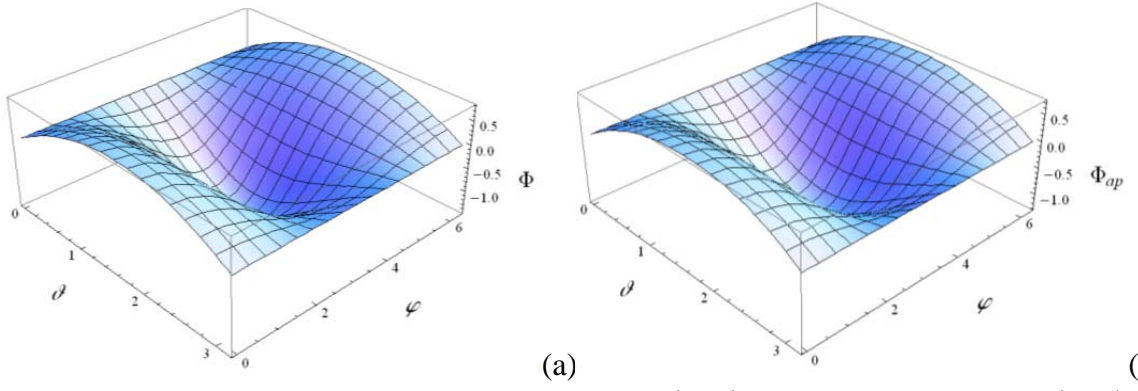

(a)                                          (b)

FIG. 3. 3D plot of the nonconservative potentials $\Phi(\vartheta,\varphi)$, Eq. (18), (a) and $\Phi_{ap}(\vartheta,\varphi)$, Eq. (19), (b) for typical values of the model parameters ($P = 0.3$, $J = 6$, $\varphi_P = 0$, and $\vartheta_P = \pi/2$).

$$B_{1,m} = \frac{4\pi b_p J}{3} Y_{1,m}^*(\vartheta_P,\varphi_P), \quad m=0,\pm 1$$

$$B_{2,0} = -\frac{8\pi^{3/2} b_P c_P J}{9\sqrt{5}} \left[ Y_{1,0}^{*2}(\vartheta_P,\varphi_P) + Y_{1,1}^*(\vartheta_P,\varphi_P) Y_{1,-1}^*(\vartheta_P,\varphi_P) \right],$$

$$B_{2,\pm 1} = -\frac{8\pi^{3/2} b_P c_P J}{3\sqrt{15}} Y_{1,0}^*(\vartheta_P,\varphi_P) Y_{1,\pm 1}^*(\vartheta_P,\varphi_P),$$

$$B_{2,\pm 2} = -\frac{8\pi^{3/2} b_P c_P J}{3\sqrt{30}} Y_{1,\pm 1}^{*2}(\vartheta_P,\varphi_P).$$

We remark that the approximation Eq. (19) accurately reproduces all features of the nonconservative potential $\Phi$, Eq. (18) (see Fig. 3) because $(c_P \mathbf{u} \cdot \mathbf{e}_P)^2 / 3 < 0.15$ for $P \leq 0.4$ ( $P \approx 0.3 \div 0.4$ are typical values for ferromagnetic metals[13]) and all $\vartheta$, $\varphi$, $\vartheta_P$, and $\varphi_P$. Moreover, in the calculation of the statistical moments, relaxation time, etc., the approximation Eq. (19) yields an accuracy better than 5% in the majority of cases.

## IV. CALCULATION OF OBSERVABLES

The general time-dependent Eq. (10) and the time independent Eq. (14), as specialized to Eqs. (17) and (19), now yields the 25-term differential-recurrence relation for the statistical moments $\langle Y_{l,m} \rangle(t)$ governing the dynamics of the magnetization [see Eq. (42) of Appendix C]. Equations (10) and (14) may be solved by extending the general matrix continued fraction methods developed in Refs. 23 and 32 to include the STT. Indeed, we can always transform the moment system, Eq. (10) constituting a multiterm *scalar* differential-recurrence relation, governing the magnetization relaxation into the *tridiagonal vector* differential-recurrence relation

$$\tau_N \dot{\mathbf{C}}_n(t) = \mathbf{Q}_n^- \mathbf{C}_{n-1}(t) + \mathbf{Q}_n \mathbf{C}_n(t) + \mathbf{Q}_n^+ \mathbf{C}_{n+1}(t). \tag{20}$$

Here $\mathbf{C}_n(t)$ are the column vectors arranged in an appropriate way from the entries $\langle Y_{l,m}\rangle(t)$ given by Eq. (11) and $\mathbf{Q}_n^{\pm}, \mathbf{Q}_n$ are matrices formed from $e_{l',m',l,m}$. The explicit equations for $\mathbf{C}_n(t)$ and $\mathbf{Q}_n^{\pm}, \mathbf{Q}_n$ for the free energy Eq. (3) are given explicitly in Appendix C. As shown in Ref. 23, Chap. 2, the exact matrix continued fraction solution of Eq. (20) for the Laplace transform of $\mathbf{C}_1(t)$ is given by

$$\tilde{\mathbf{C}}_1(s) = \tau_N \mathbf{\Delta}_1(s)\left\{\mathbf{C}_1(0) + \sum_{n=2}^{\infty}\left[\prod_{k=2}^{n}\mathbf{Q}_{k-1}^{+}\mathbf{\Delta}_k(s)\right]\mathbf{C}_n(0)\right\}, \tag{21}$$

where $\tilde{\mathbf{C}}_1(s) = \int_0^{\infty}\mathbf{C}_1(t)e^{-st}dt$, $\mathbf{\Delta}_n(s)$ is the matrix continued fraction defined by the recurrence relation

$$\mathbf{\Delta}_n(s) = \left[\tau_N s\mathbf{I} - \mathbf{Q}_n - \mathbf{Q}_n^{+}\mathbf{\Delta}_{n+1}(s)\mathbf{Q}_{n+1}^{-}\right]^{-1}, \tag{22}$$

and $\mathbf{I}$ is the unit matrix. Having determined the column vectors $\tilde{\mathbf{C}}_1(s)$, $\tilde{\mathbf{C}}_2(s)$, ... as described in Refs. 23 and 32, we then have the relevant observables. In a similar way, we also have the smallest nonvanishing eigenvalue (yielding the reversal time) from the matrix equation[3,23]

$$\lambda\tau_N\mathbf{I} = \mathbf{S} \tag{23}$$

where the matrix $\mathbf{S}$ is defined via the matrix continued fractions as

$$\mathbf{S} = -\left[\mathbf{Q}_1 + \mathbf{Q}_1^{+}\mathbf{\Delta}_2(0)\mathbf{Q}_2^{-}\right]\left[\mathbf{I} - \mathbf{Q}_1^{+}\mathbf{\Delta}_2'(0)\mathbf{Q}_2^{-}\right]^{-1} \tag{24}$$

and the prime designates the derivative of $\mathbf{\Delta}_2(s)$ with respect to $\tau_N s$ (see Ref. 23, Chap. 2, Sec. 2.11.2). Thus $\lambda_1$ is the smallest nonvanishing eigenvalue of $\mathbf{S}$.

Now in order to calculate the stationary characteristics (distribution function, magnetization, etc.) we may replace $\langle Y_{l,m}\rangle(t)$ in Eq. (42) of Appendix C by $\langle Y_{l,m}\rangle_0$ and set the time derivative equal to zero. Then by extending the general matrix continued fraction methods developed in Refs. 23 and 32 to include the STT, we have

$$\begin{pmatrix} \langle Y_{2n,-2n}\rangle_0 \\ \langle Y_{2n,-2n+1}\rangle_0 \\ \vdots \\ \langle Y_{2n,2n}\rangle_0 \\ \langle Y_{2n-1,-2n+1}\rangle_0 \\ \langle Y_{2n-1,-2n+2}\rangle_0 \\ \vdots \\ \langle Y_{2n-1,2n-1}\rangle_0 \end{pmatrix} = \frac{1}{\sqrt{4\pi}}\mathbf{\Delta}_n(0)\mathbf{Q}_n^{-}\mathbf{\Delta}_{n-1}(0)\mathbf{Q}_{n-1}^{-}\cdots\mathbf{\Delta}_1(0)\mathbf{Q}_1^{-}, \quad (n=1,2,....). \tag{25}$$

The out-of-equilibrium or time-independent stationary distribution $W_0(\vartheta,\varphi)$ is thus rendered via the Fourier series

$$W_0(\vartheta,\varphi) = \sum_{l=0}^{\infty}\sum_{m=-l}^{l} \langle Y_{l,m}\rangle_0 Y_{l,m}^*(\vartheta,\varphi). \tag{26}$$

Next, by analogy with the Boltzmann distribution, since we expect the stationary distribution to be formally similar to it,[13] we may define the *effective potential* $V_{ef}$ via

$$V_{ef}(\vartheta,\varphi) = -\ln[W_0(\vartheta,\varphi)]. \tag{27}$$

Furthermore, having determined $\langle Y_{l,m}\rangle_0$, we have both the stationary magnetization in the X direction $\langle u_X\rangle_0$ and the corresponding susceptibility $\propto \langle u_X^2\rangle_0 - \langle u_X\rangle_0^2$, viz.,

$$\langle u_X\rangle_0 = \langle \sin\vartheta\cos\varphi\rangle_0 = \sqrt{\frac{2\pi}{3}}\left(\langle Y_{1,-1}\rangle_0 - \langle Y_{1,1}\rangle_0\right), \tag{28}$$

$$\begin{aligned}\langle u_X^2\rangle_0 - \langle u_X\rangle_0^2 &= \langle \sin^2\vartheta\cos^2\varphi\rangle_0 - \langle \sin\vartheta\cos\varphi\rangle_0^2 \\ &= \sqrt{\frac{2\pi}{15}}\left(\langle Y_{2,2}\rangle_0 + \langle Y_{2,-2}\rangle_0\right) - \sqrt{\frac{4\pi}{45}}\langle Y_{2,0}\rangle_0 - \frac{2\pi}{3}\left(\langle Y_{1,-1}\rangle_0 - \langle Y_{1,1}\rangle_0\right)^2 + \frac{1}{3}.\end{aligned} \tag{29}$$

We remark that in some ranges of the model parameters, e.g., for very low damping $\alpha$ < 0.001, and/or very high potential barriers, $\Delta V > 100$, the continued fraction method may not be applicable[18,31] because the matrices involved become ill-conditioned, meaning that numerical inversions are no longer possible.

## V. RESULTS

Throughout the calculations the anisotropy and spin-polarization parameters will be taken as $D_{\parallel} = 0.034$, $\delta = 20$, and $P = 0.3$ just as in Ref. 13. Moreover, the applied field $\mathbf{H}_0$ and the unit vector $\mathbf{e}_P$ identifying the magnetization direction in the fixed layer are taken to lie in the equatorial or XY plane, i.e., $\vartheta_P = \pi/2$ and $\vartheta_\xi = \pi/2$. Thus the orientations of $\mathbf{H}_0$ and $\mathbf{e}_P$ in the XY plane are entirely determined by the azimuthal angles $\varphi_\xi$ and $\varphi_P$, respectively. The values $\varphi_\xi = \varphi_P = 0$ correspond to the particular configuration whereby both $\mathbf{H}_0$ and $\mathbf{e}_P$ are directed along the easy (X-)axis. For $\gamma = 2.2\cdot 10^5$ m A$^{-1}$s$^{-1}$, $T = 300$ K, $v \sim 10^{-24}$ m$^3$, $M_S \approx 1.4\cdot 10^6$ A m$^{-1}$ (cobalt), $J_e \approx 10^7$ A cm$^{-2}$, $J_p \approx 10^9$ A cm$^{-2}$, and $\alpha = 0.02$, we have the following estimates for the principal model parameters

$$J \approx 5.9, \ \sigma \approx 20.2, \ \tau_N \approx 4.8\cdot 10^{-8}\,\text{s}.$$

Moreover, instead of the free diffusion time $\tau_N$, it will be more convenient to use as the normalizing time $\tau_0 = \tau_N / [\sigma(\alpha + \alpha^{-1})] = (2\gamma M_S D_\parallel)^{-1}$. The above numerical values yield $\tau_0 \approx 4.8 \cdot 10^{-11}$ s.

Once we have determined the time-independent stationary distribution $W_0(\vartheta, \varphi)$ via the Fourier series, Eq. (26), we have the effective potential $V_{ef}$ from Eq. (27). A typical example of such calculations is shown in Figs. 4 and 5. The effective potential comprises a double-well structure with non-equivalent wells. This energyscape (Fig. 4), as expected on intuitive grounds, strongly depends on damping, external magnetic field magnitude and orientation, and spin-polarized current. In particular, by varying the magnitude of the spin-polarized current, damping, etc., one may alter substantially the effective barriers and thus the reversal time (cf. Fig. 5). The stationary averages $\langle u_X \rangle_0$ and $\langle u_X^2 \rangle_0 - \langle u_X \rangle_0^2$ are calculated from Eqs. (28) and (29), respectively. In Fig. 6, we show the consequent dependence of $\langle u_X \rangle_0$ and $\langle u_X^2 \rangle_0 - \langle u_X \rangle_0^2$ on the spin-polarized current via a family of curves with the spin current as the independent variable, for various values of the damping, external field magnitude ($h$) and orientation ($\varphi_\xi$), and the magnetization direction in the fixed layer ($\varphi_P$), which are all regarded as parameters. In contrast, in Fig. 7 we illustrate the dependence of the magnetisation and the susceptibility on the external field parameter $h$ via a family of curves for various values of the current $J$, the external field orientation and magnetization direction in the fixed layer ($\varphi_\xi$ and $\varphi_P$), and inverse temperature parameter $\sigma$. Clearly the switching current $J_{SW}$ (i.e., the current when $\langle u_X \rangle_0$ changes sign corresponding to reversal of the direction of precession or, equivalently, when $\langle u_X^2 \rangle_0 - \langle u_X \rangle_0^2$ attains its maximum and subsequently vanishes) strongly depend on the model parameters $\alpha$, $J$, $h$, $\sigma$, $\delta$, $\varphi_\xi$, and $\varphi_P$. In particular, as both the damping $\alpha$ and external field parameters $h$ increase the value of $J_{SW}$ rapidly increases [see Fig. 6(a), 6(b), and 7(a)]. Moreover, $J_{SW}$ may also vary significantly with both the orientation of the external field and the direction of the magnetization of the fixed layer (see Fig. 6c and 6d). The half-width of $\langle u_X^2 \rangle_0 - \langle u_X \rangle_0^2$ and the onset of the slope in $\langle u_X \rangle_0$ largely depend on the damping parameter $\alpha$ [Fig. 6(a)] and the inverse temperature parameter $\sigma$ [Fig. 7(d)]: higher values of $\sigma$ (lower temperatures) and smaller values of $\alpha$ correspond to a narrower half-width and a more rapid onset of the slope.

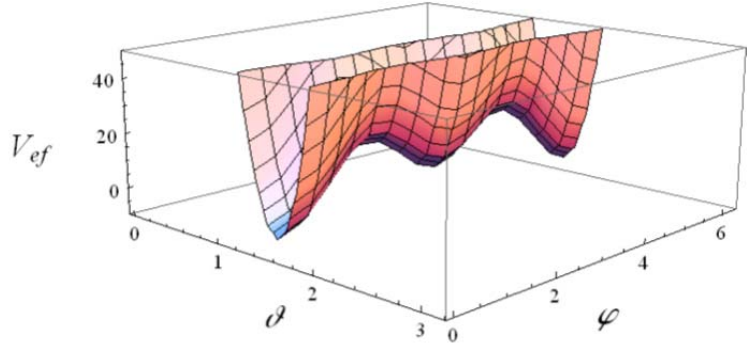

FIG. 4. (a) 3D plot of the effective potential $V_{ef}(\vartheta,\varphi)$, Eq. (27), in the vicinity of the minima for $J = 6$, $\alpha = 0.02$, $h = 0.15$, $\sigma = 20$, $\delta = 20$, $\varphi_P = 0$, and $\varphi_\xi = 0$.

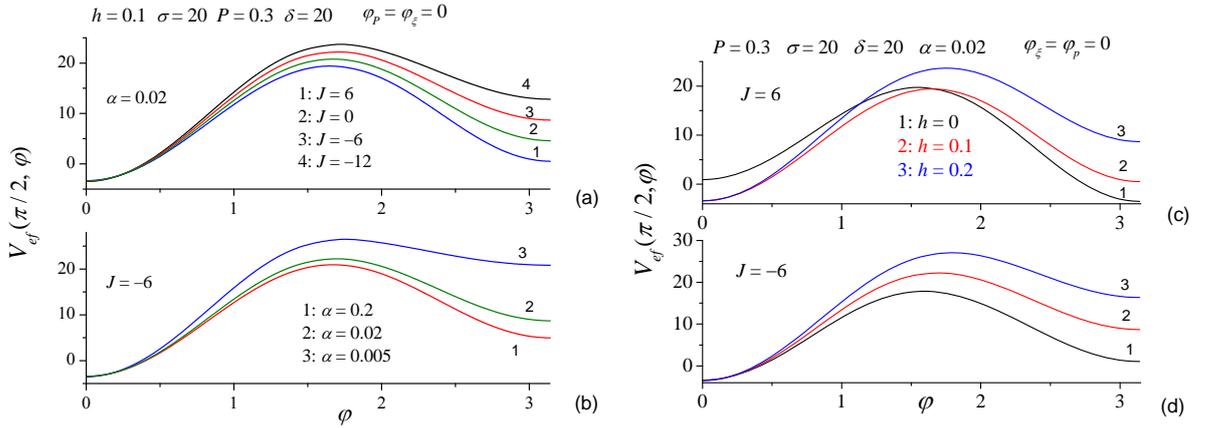

FIG. 5. $V_{ef}(\pi/2,\varphi)$ vs the azimuthal angle $\varphi$ for various spin polarized current parameter $J = 6, 0, -6, -12$ and $\alpha = 0.02$ (a); for various damping $\alpha = 0.2, 0.02, 0.005$ and $J = -6$ (b); for various external field parameter $h = 0.0, 0.1, 0.2$ and $J = 6$ (c) and $J = -6$ (d) ($\alpha = 0.02$, $h = 0.15$, $\sigma = 20$, $\delta = 20$, $\varphi_P = 0$, and $\varphi_\xi = 0$).

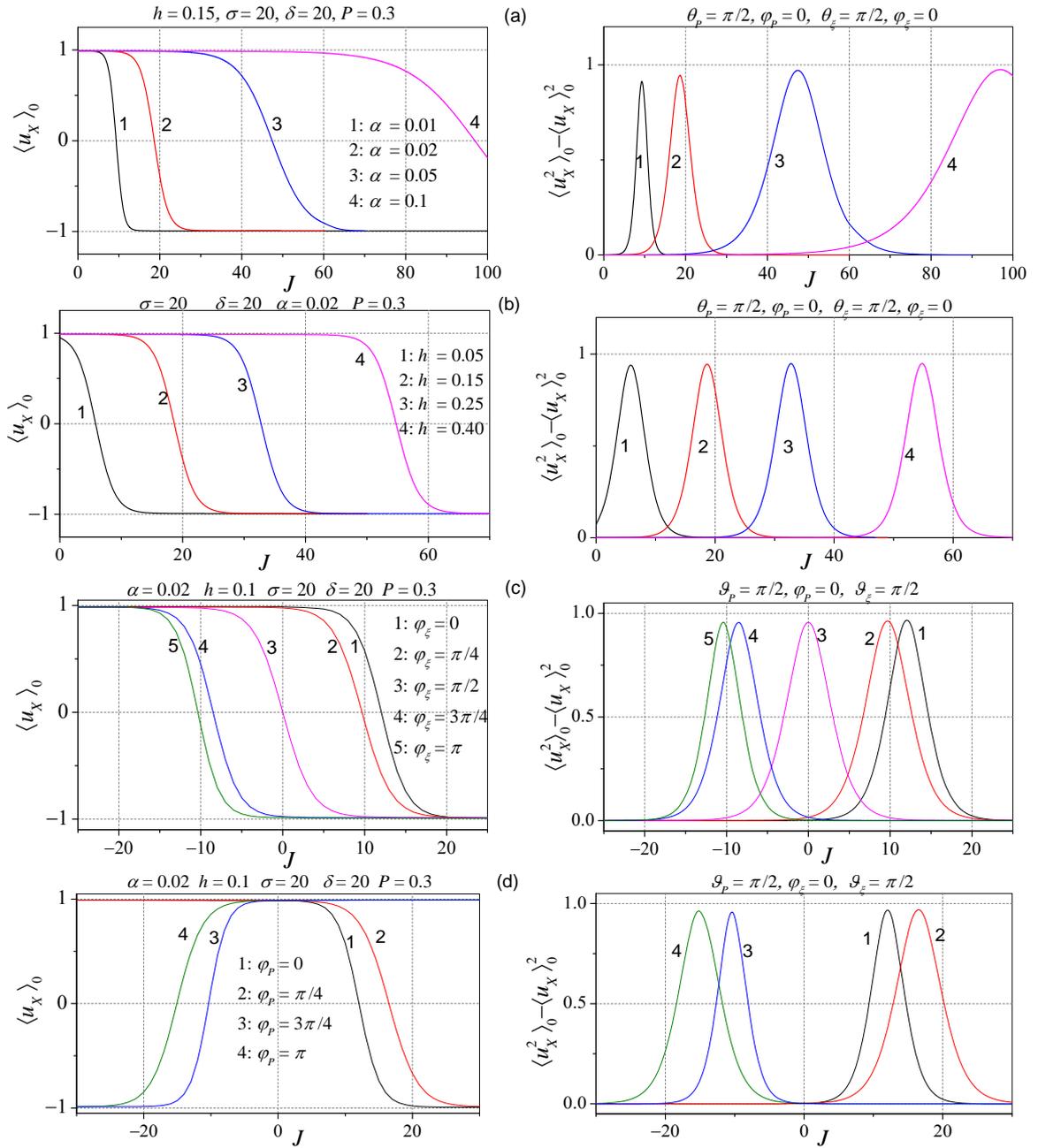

FIG. 6. $\langle u_X \rangle_0$ and $\langle u_X^2 \rangle_0 - \langle u_X \rangle_0^2$ vs the dimensionless current parameter $J$ for various values of damping $\alpha$ (a), external field parameter $h$ (b), external field orientation in the free layer $\varphi_\xi$ (c), and spin-polarization orientation $\varphi_P$ (d).

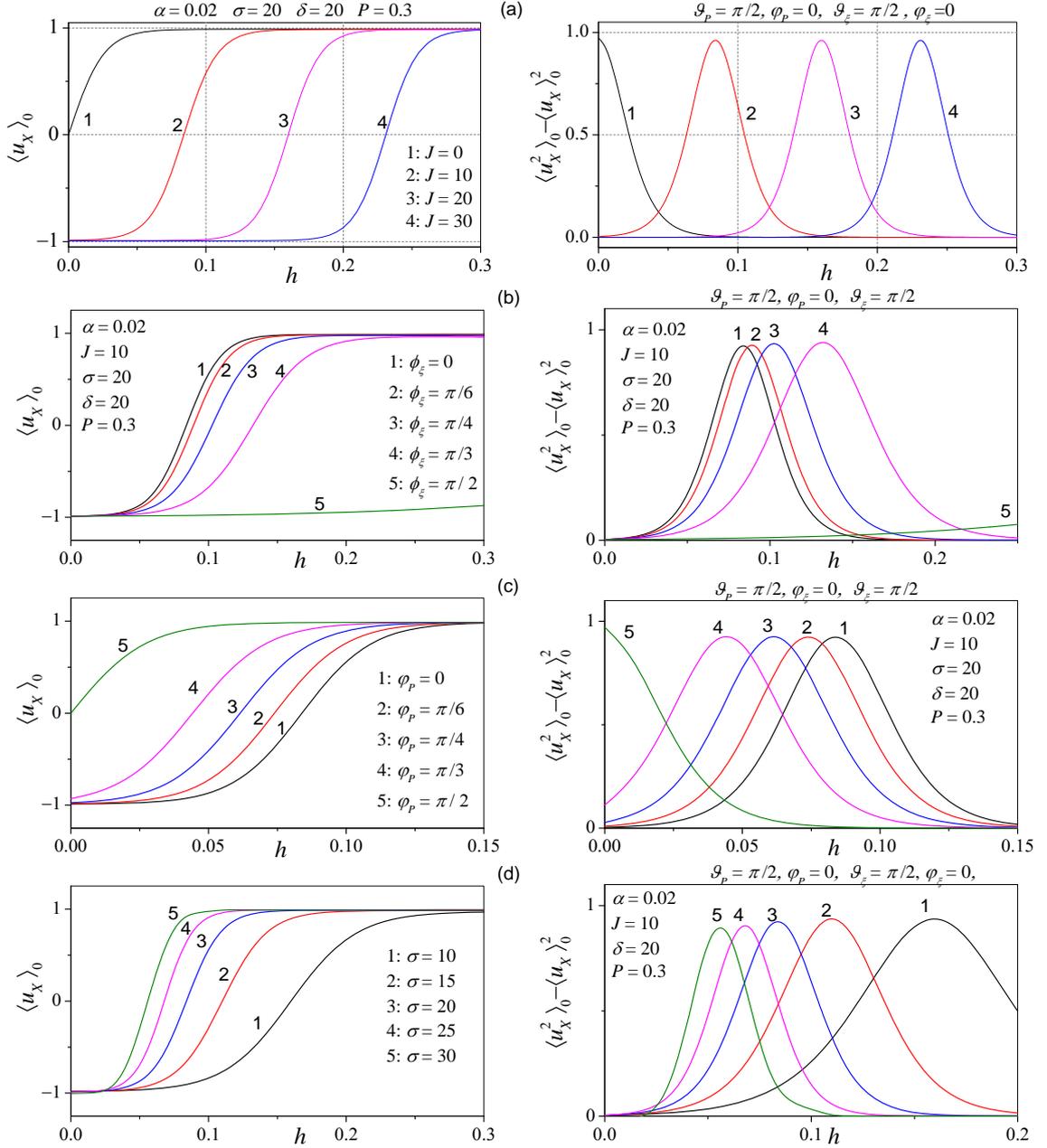

FIG. 7. $\langle u_X \rangle_0$ and $\langle u_X^2 \rangle_0 - \langle u_X \rangle_0^2$ vs the external field parameter $h$ for various values of the current parameter $J$ (a), the external field orientation $\varphi_\xi$ (b), the spin-polarization orientation $\varphi_P$ (c), and the inverse temperature parameter $\sigma$ (d).

Moreover, the smallest nonvanishing eigenvalue $\lambda_1$ of the Fokker-Planck operator (inverse of the reversal time of the direction of precession, i.e., that of the $X$ component of $\mathbf{M}$) may be evaluated from the secular Eq. (23). By calculating $\lambda_1$, we also have the dependence of the magnetization reversal time $\tau = 1/\lambda_1$ on the spin-polarized current, anisotropy parameters, damping, external field magnitude and orientation in the free layer, and the magnetization direction in the fixed layer. In general, a pronounced dependence of the reversal time on these model parameters exists. Examples are shown in Figs. 8–11. Figure 8 illustrates the damping dependence of $\tau$ for various values of the current $J$ while Fig. 9 shows the temperature dependence of $\tau$ for various values of the current ($J$) and external field ($h$) parameters. We have also shown for comparison in these figures the reversal time calculated for $J = 0$ via escape rate theory for biaxial anisotropy which is given by[35]

$$\tau = \frac{A(\alpha S_1 + \alpha S_2)}{2(\Gamma_1^{IHD} + \Gamma_2^{IHD})A(\alpha S_1)A(\alpha S_2)}. \quad (30)$$

where $A(z)$ is called the depopulation factor, viz.

$$A(z) = \exp\left(\frac{1}{2\pi}\int_{-\infty}^{\infty}\ln\left[1 - e^{-z\left(\lambda^2 + 1/4\right)}\right]\frac{d\lambda}{\lambda^2 + 1/4}\right), \quad (31)$$

$\Gamma_2^{IHD}$ is the escape rate from the shallower well 2 to a deeper well 1 given by

$$\Gamma_2^{IHD}(h) = \frac{e^{-(1-h)^2\sigma}}{2\pi\tau_0(\alpha + \alpha^{-1})}\sqrt{\frac{1-h+\delta}{\delta(1+h)}}\left[1 - h^2 - \delta + \sqrt{(1-h^2+\delta)^2 + 4\delta(1-h^2)\alpha^{-2}}\right], \quad (32)$$

$\Gamma_1(h) = \Gamma_2(-h)$, and the dimensionless actions $S_1$ and $S_2$ are given by

$$S_{1,2} = \frac{4\delta\sigma(1-h^2+\delta)}{(1+\delta)^{3/2}}\left\{\left[(1+\delta^{-1})(1-h^2)\right]^{1/2} + h\arctan\left[h\left[(1-h^2)(1+\delta^{-1})\right]^{-1/2}\right] \pm \frac{h\pi}{2}\right\}. \quad (33)$$

Clearly by altering $J$, the ensuing variation of $\tau$ may be as much as several orders of magnitude for very low damping, $\alpha \ll 1$ (Fig. 8). Furthermore, $\tau$ may greatly exceed or, on the other hand, be much less than the value pertaining to $J = 0$. Moreover, the increase or decrease in $\tau$ is entirely governed by the direction of the current, i.e., by the sign of the parameter $J$ as expected.[3] The temperature dependence of $\tau$ can be understood via the *effective potential* $V_{ef}(\vartheta,\varphi)$, Eq. (27). Clearly, at high barriers, $\sigma > 5$, the *temperature dependence* of $\tau$ has the customary Arrhenius behavior $\tau \sim e^{v\Delta V_{ef}/(kT)}$, i.e., exponentially increasing with decreasing temperature. The slope of $\tau(T^{-1})$ markedly depends on $J$, $h$, $\alpha$, etc. because the barrier height $\Delta V_{ef}$ of the shallow well is strongly influenced by those parameters (see Fig. 5). In particular, we observe that the slope of $\tau(T^{-1})$ significantly decreases with increasing $h$ [Fig. 9(b)] due to a decrease of the barrier height $\Delta V_{ef}$ due to the action of the external field [see Fig. 5(c) and 5(d)]. At low barriers, the behavior of $\tau(T^{-1})$ may deviate considerably from Arrhenius behavior. Figure 10 illustrates the dependence of $\tau$ on the current parameter $J$ for various values of the external field parameter $h$. Clearly, as $J$ increases from negative values, $\tau$ exponentially increases attaining a maximum at a critical value of the spin-polarized current and then smoothly switches to exponential decrease as $J$ is further increased through positive values. Such a dependence of $\tau$ on the applied current implies that both kinds of scaling for the switching time suggested in the literature,[24-26,28] namely, $\tau \propto e^{-C|J_{SW}-J|}$ and $\tau \propto e^{-C(J_{SW}-J)^2}$ may be realized for $|J_{SW}-J| > 10$ and $|J_{SW}-J| < 10$, respectively (where $C$ and $J_{SW}$ are parameters depending, in general, on $h$, $\sigma$, $\delta$, $\alpha$, etc.). Figure 11 exemplifies the pronounced dependence of $\tau$ on the azimuthal angles of the applied field $\varphi_\xi$ and the magnetization direction in the fixed layer $\varphi_P$ for various values of $J$ which may comprise several orders of magnitude [note that $\tau(\varphi_\xi) = \tau(2\pi - \varphi_\xi)$ and $\tau(\varphi_P) = \tau(2\pi - \varphi_P)$]. Invariably strong STT effects on the magnetization reversal exist only for low damping, $\alpha \leq 0.1$, because the magnitude of the STT effects in the magnetization reversal is governed by the ratio $J/\alpha$.[13] Here, the variation of $\tau$ with $J$ may be of several orders of magnitude. For $\alpha \geq 1$, however, the STT term in Eq. (1) does not influence the reversal process at all because it is negligible compared to the damping and random contributions.

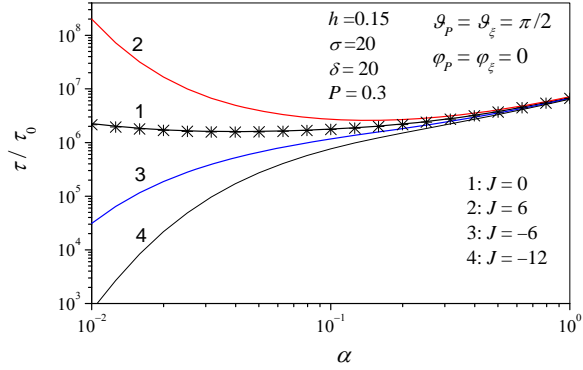

FIG. 8. Reversal time $\tau/\tau_0$ vs the damping parameter $\alpha$ for various values of the current $J$ (solid lines). Asterisks: escape rate formula, Eq.(30).

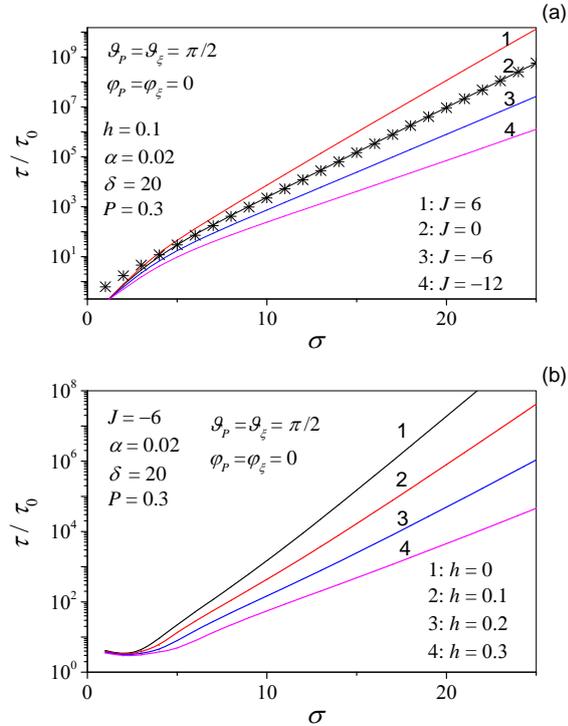

FIG. 9. $\tau/\tau_0$ vs the inverse temperature parameter $\sigma \sim T^{-1}$ for various values of the current $J$ (a) and the external field $h$ (b). Asterisks: escape rate formula, Eq.(30).

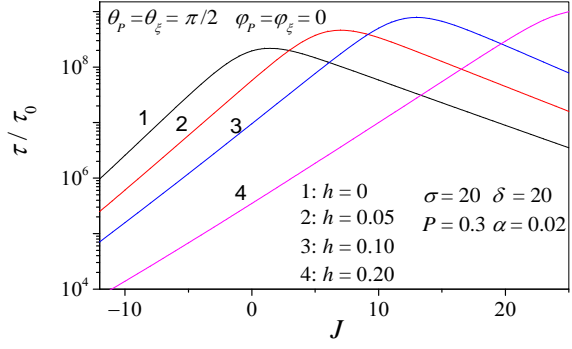

FIG. 10. $\tau/\tau_0$ vs the spin-polarized current parameter $J$ for various values of the external field $h = 0, 0.05, 0.1, 0.2$.

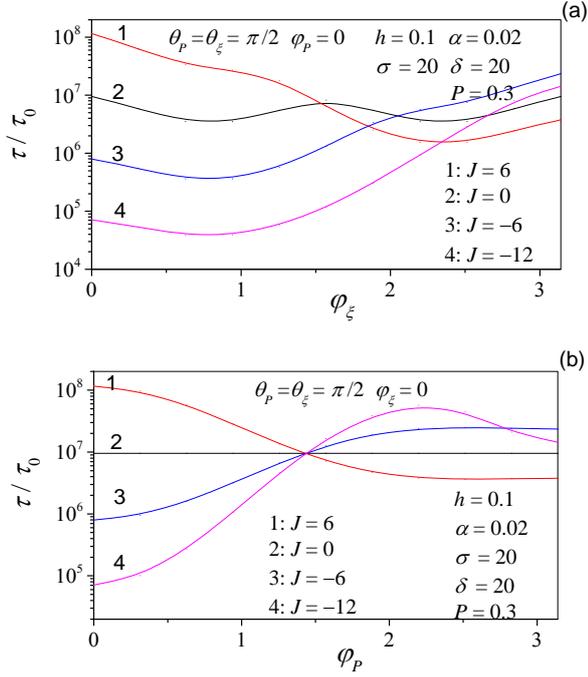

FIG. 11. $\tau/\tau_0$ vs the azimuthal angles $\varphi_\xi$ (a) and $\varphi_P$ (b) for $J = 6, 0, -6, -12$.

STT effects in the thermally assisted magnetization reversal have been treated via the evolution equation for the statistical moments yielded by the Langevin equation rendering stationary and nonstationary characteristics for wide ranges of temperature, damping, external magnetic field, and spin-polarized current. Variation of the latter may alter the reversal time by several orders of magnitude concurring with experimental results.[11] The virtue of our numerically exact solutions of the recurrence relations for the relevant statistical moments is that they hold for the most comprehensive formulation of the generic nanopillar model (Fig. 1), i.e., for *arbitrary directions* of the external field and spin polarization and for *arbitrary free energy density,* yielding the STT switching characteristics under conditions otherwise inaccessible. Thus our results may serve both as a basis for theoretical investigations and interpretation of a broad range of STT experiments. Additionally, they are essential for the future development of both escape rate theory and stochastic dynamics simulations of the magnetization reversal time in STT systems, representing rigorous benchmark solutions with which calculations of that time by any other method must comply (the procedure being entirely analogous to that used to validate such complementary approaches to fine particle magnetization by exactly calculating the reversal time via the smallest nonvanishing eigenvalue of the FPE for the Néel-Brown model[3,36]). Finally, we believe that the moment method may be useful in related problems such as magnetization reversal of STT devices driven by ac external fields and currents, etc.[37]

## ACKNOWLEDGEMENTS

We thank P.M. Déjardin for helpful conversations. One of us, D.B., acknowledges the SimSci Structured Ph.D. Programme at the University College Dublin for financial support. SimSci is funded under the Programme for Research in Third-level Institutions and co-funded under the European Regional Development Fund. All calculations were performed by the Lonsdale cluster maintained by the Trinity Centre for High Performance Computing. This Cluster was funded through grants from the Science Foundation Ireland.

## APPENDIX A: REDUCTION OF EQ. (1) TO THE LANDAU-LIFSHITZ FORM

In order to obtain an explicit equation for $\dot{\mathbf{u}}$, we rewrite the implicit Eq. (1) in the Landau-Lifshitz form. Transposing the $\dot{\mathbf{u}}$ term, we have

$$\dot{\mathbf{u}} + \alpha[\dot{\mathbf{u}} \times \mathbf{u}] = -\gamma\left[\mathbf{u} \times \left(\mathbf{H}_{ef} + \dot{\mathbf{u}}\big|_{ST} + \mathbf{h}\right)\right]. \tag{34}$$

On cross-multiplying vectorially by $\mathbf{u}$ in Eq. (34) and using the triple vector product formula

$$\left[[\dot{\mathbf{u}} \times \mathbf{u}] \times \mathbf{u}\right] = -\dot{\mathbf{u}} + \mathbf{u}(\mathbf{u} \cdot \dot{\mathbf{u}}), \tag{35}$$

we obtain since $(\mathbf{u} \cdot \dot{\mathbf{u}}) = 0$

$$[\dot{\mathbf{u}} \times \mathbf{u}] = \alpha\dot{\mathbf{u}} - \gamma\left[\left[\mathbf{u} \times \left(\mathbf{H}_{ef} + \dot{\mathbf{u}}\big|_{ST} + \mathbf{h}\right)\right] \times \mathbf{u}\right]. \tag{36}$$

Substituting Eq. (36) into Eq. (34) yields the explicit form

$$(1+\alpha^2)\dot{\mathbf{u}} = -\gamma\left[\mathbf{u} \times \left(\mathbf{H}_{ef} + \dot{\mathbf{u}}\big|_{ST} + \mathbf{h}\right)\right] + \gamma\alpha\left[\left[\mathbf{u} \times \left(\mathbf{H}_{ef} + \dot{\mathbf{u}}\big|_{ST} + \mathbf{h}\right)\right] \times \mathbf{u}\right]$$

or equivalently, using Eq. (6),

$$\begin{aligned}\dot{\mathbf{u}} &= \frac{v}{2kT\alpha\tau_N}\left[\mathbf{u} \times \left(\frac{\partial V}{\partial \mathbf{u}} + \left[\mathbf{u} \times \frac{\partial \Phi}{\partial \mathbf{u}}\right] - \mu_0 M_S \mathbf{h}\right)\right] \\ &+ \frac{v}{2kT\tau_N}\left[\mathbf{u} \times \left[\mathbf{u} \times \left(\frac{\partial V}{\partial \mathbf{u}} + \left[\mathbf{u} \times \frac{\partial \Phi}{\partial \mathbf{u}}\right] - \mu_0 M_S \mathbf{h}\right)\right]\right]\end{aligned} \tag{37}$$

which reduces to the more easily visualized Eq. (6) because

$$\left[\mathbf{u}\times\left[\mathbf{u}\times\left[\mathbf{u}\times\frac{\partial\Phi}{\partial\mathbf{u}}\right]\right]\right]=-\left[\mathbf{u}\times\frac{\partial\Phi}{\partial\mathbf{u}}\right].$$

In the spherical polar coordinate system shown in Fig. 1, one has

$$\mathbf{u}=(1,0,0),\ \dot{\mathbf{u}}=\left(0,\dot{\vartheta},\dot{\varphi}\sin\vartheta\right),$$

$$\frac{\partial V}{\partial\mathbf{u}}=\left(0,\frac{\partial V}{\partial\vartheta},\frac{1}{\sin\vartheta}\frac{\partial V}{\partial\varphi}\right),$$

$$\left[\mathbf{u}\times\frac{\partial\Phi}{\partial\mathbf{u}}\right]=\left(0,\frac{-1}{\sin\vartheta}\frac{\partial\Phi}{\partial\varphi},\frac{\partial\Phi}{\partial\vartheta}\right).$$

## APPENDIX B: DERIVATION OF THE DIFFERENTIAL-RECURRENCE RELATION FOR STATISTICAL MOMENTS, EQ. (10)

Owing to Eqs. (7) and (8), we can write the Langevin equation (9) in vector notation as

$$\dot{Y}_{l,m}=-\frac{v\mu_0 M_S}{2kT\tau_N}\left(\mathbf{h}\cdot\{[\mathbf{u}\times\nabla Y_{l,m}]-\alpha^{-1}\nabla Y_{l,m}\}\right)$$
$$-\frac{v}{2kT\tau_N}\left(\{\nabla(V+\alpha^{-1}\Phi)-[\mathbf{u}\times\nabla(\alpha^{-1}V-\Phi)]\}\cdot\nabla Y_{l,m}\right). \quad (38)$$

Here $\nabla$ denotes the orientation space gradient operator defined as

$$\nabla=\left[\mathbf{u}\times\frac{\partial}{\partial\mathbf{u}}\right].$$

On averaging the Langevin equation (38) with multiplicative noise as explained in detail in Ref. 32 and Sec. 9.2 of Ref. 23, we have after some algebra the evolution equation of the statistical moment $\langle Y_{l,m}\rangle(t)$, viz.,

$$\frac{d}{dt}\langle Y_{l,m}\rangle=\frac{1}{2\tau_N}\left\langle\Delta Y_{l,m}+\frac{v}{2kT}\left\{(V+\alpha^{-1}\Phi)\Delta Y_{l,m}+Y_{l,m}\Delta(V+\alpha^{-1}\Phi)-\Delta((V+\alpha^{-1}\Phi)Y_{l,m})\right.\right.$$
$$\left.\left.-\frac{1}{\sin\vartheta}\left[\frac{\partial(\alpha^{-1}V-\Phi)}{\partial\varphi}\frac{\partial Y_{l,m}}{\partial\vartheta}-\frac{\partial(\alpha^{-1}V-\Phi)}{\partial\vartheta}\frac{\partial Y_{l,m}}{\partial\varphi}\right]\right\}\right\rangle \quad (39)$$

where the operator $\Delta=\nabla^2$ is the angular part of the Laplacian, viz.,

$$\Delta=\frac{1}{\sin\vartheta}\frac{\partial}{\partial\vartheta}\left(\sin\vartheta\frac{\partial}{\partial\vartheta}\right)+\frac{1}{\sin^2\vartheta}\frac{\partial^2}{\partial\varphi^2}.$$

Here we have used[23]

$$\frac{v\mu_0 M_S}{kT}\overline{\left(\mathbf{h}\cdot\{[\mathbf{u}\times\nabla Y_{l,m}]-\alpha^{-1}\nabla Y_{l,m}\}\right)}=-\Delta Y_{l,m}$$

and that for any function $F(\vartheta,\varphi,t)$[23]

$$2\nabla F\cdot\nabla Y_{l,m}=\Delta(FY_{l,m})-F\Delta Y_{l,m}-Y_{l,m}\Delta F.$$

We now indicate using the theory of angular momentum[34] how Eq. (39) may be written as a differential-recurrence relation for the statistical moments. This is accomplished by reducing (details in Refs. 32 and 23, chapters 7 and 9) the terms inside the angular braces on the right-hand side of Eq. (39) to the calculation of the Fourier coefficients in the expansion of a product of spherical harmonics as a sum of spherical harmonics. We begin by expressing the terms within the angular braces on the right-hand side of Eq. (39) as functions of the angular momentum operators $\hat{L}^2$, $\hat{L}_z$, and $\hat{L}_\pm$ defined as[34]

$$\hat{L}^2 = -\Delta,\ \hat{L}_z = -i\frac{\partial}{\partial \varphi},\ \hat{L}_\pm = \pm e^{\pm i\varphi}\frac{\partial}{\partial \vartheta} + i\cot\vartheta\, e^{\pm i\varphi}\frac{\partial}{\partial \varphi}. \tag{40}$$

Thus we have from Eqs. (39) and (40) the evolution equation

$$\frac{d}{dt}\langle Y_{l,m}\rangle = \frac{1}{2\tau_N}\Bigg\langle -\hat{L}^2 Y_{l,m} + \frac{\beta}{2}\bigg[\hat{L}^2\bigg(\bigg(V+\frac{1}{\alpha}\Phi\bigg)Y_{l,m}\bigg) - \bigg(V+\frac{1}{\alpha}\Phi\bigg)\hat{L}^2 Y_{l,m} - Y_{l,m}\hat{L}^2\bigg(V+\frac{1}{\alpha}\Phi\bigg)\bigg]$$
$$+\frac{i}{2\alpha}\sqrt{\frac{3}{2\pi}}\Big\{Y_{1,1}^{-1}\Big[\big(\hat{L}_z(V_+ -\alpha\Phi_+)\big)\big(\hat{L}_+ Y_{l,m}\big) - \big(\hat{L}_+(V_+ -\alpha\Phi_+)\big)\big(\hat{L}_z Y_{l,m}\big)\Big] \tag{41}$$
$$+Y_{1,-1}^{-1}\Big[\big(\hat{L}_z(V_- -\alpha\Phi_-)\big)\big(\hat{L}_- Y_{l,m}\big) - \big(\hat{L}_-(V_- -\alpha\Phi_-)\big)\big(\hat{L}_z Y_{l,m}\big)\Big]\Big\}\Bigg\rangle,$$

where we have used the following representations for the expansions of $vV/(kT) = V_- + V_+$ and $v\Phi/(kT) = \Phi_- + \Phi_+$ in terms of spherical harmonics, viz.,

$$V_- = \sum_{r=1}^{\infty}\sum_{s=-r}^{-1} A_{r,s}Y_{r,s},\ V_+ = \sum_{r=1}^{\infty}\sum_{s=0}^{r} A_{r,s}Y_{r,s},$$

$$\Phi_- = \sum_{r=1}^{\infty}\sum_{s=-r}^{-1} B_{r,s}Y_{r,s},\ \Phi_+ = \sum_{r=1}^{\infty}\sum_{s=0}^{r} B_{r,s}Y_{r,s}.$$

Then (see Refs. 23 and 32 for details) the right-hand side of Eq. (41) may ultimately be written as a linear combination of averages of spherical harmonics, i.e., Eq. (10), because the action of the operators $\hat{L}_z$, $\hat{L}_\pm$, $\hat{L}^2$ on $Y_{l,m}$ is[34]

$$\hat{L}_z Y_{l,m} = mY_{l,m},\ \hat{L}^2 Y_{l,m} = l(l+1)Y_{l,m},\ \hat{L}_\pm Y_{l,m} = \sqrt{l(l+1)-m(m\pm 1)}Y_{l,m\pm 1},$$

and products of spherical harmonics may always be reduced to a sum of spherical harmonics using the Clebsch-Gordan series, viz.,[34]

$$Y_{l,m}Y_{l_1,m_1} = \sqrt{\frac{(2l+1)(2l_1+1)}{4\pi}}\sum_{l_2=|l-l_1|}^{l+l_1}\frac{C_{l,0,l_1,0}^{l_2,0}C_{l,m,l_1,m_1}^{l_2,m+m_1}}{\sqrt{2l_2+1}}Y_{l_2,m+m_1}.$$

## APPENDIX C: EXPLICIT FORM OF $C_n(t)$, $Q_n^\pm$, $Q_n$, AND $e_{l,m,l',m'}$

The general Eq. (10) as specialized to Eqs. (17) and (19) yields the 25-term differential-recurrence equation for the statistical moments $c_{l,m}(t) = \langle Y_{l,m}\rangle(t)$, viz.,

$$\tau_N \frac{d}{dt} c_{n,m}(t) = v_{n,m}^{--} c_{n-2,m-2}(t) + v_{n,m}^{-} c_{n-2,m-1}(t) + v_{n,m} c_{n-2,m}(t) + v_{n,m}^{+} c_{n-2,m+1}(t) + v_{n,m}^{++} c_{n-2,m+2}(t)$$
$$+ w_{n,m}^{--} c_{n-1,m-2}(t) + w_{n,m}^{-} c_{n-1,m-1}(t) + w_{n,m} c_{n-1,m}(t) + w_{n,m}^{+} c_{n-1,m+1}(t) + w_{n,m}^{++} c_{n-1,m+2}(t)$$
$$+ x_{n,m}^{--} c_{n,m-2}(t) + x_{n,m}^{-} c_{n,m-1}(t) + x_{n,m} c_{n,m}(t) + x_{n,m}^{+} c_{n,m+1}(t) + x_{n,m}^{++} c_{n,m+2}(t) \quad (42)$$
$$+ y_{n,m}^{--} c_{n+1,m-2} + y_{n,m}^{-} c_{n+1,m-1}(t) + y_{n,m} c_{n+1,m}(t) + y_{n,m}^{+} c_{n+1,m+1}(t) + y_{n,m}^{++} c_{n+1,m+2}(t)$$
$$+ z_{n,m}^{--} c_{n+2,m-2}(t) + z_{n,m}^{-} c_{n+2,m-1}(t) + z_{n,m} c_{n+2,m}(t) + z_{n,m}^{+} c_{n+2,m+1}(t) + z_{n,m}^{++} c_{n+2,m+2}(t),$$

where the coefficients $x_{n,m}$, etc. are given by

$$x_{n,m} = -\frac{n(n+1)}{2} + i\frac{m\sigma h\gamma_3}{\alpha} + i\sqrt{\frac{\pi}{3}} mb_P J Y_{10}^*(\vartheta_P, \varphi_P)$$
$$+ \frac{n(n+1)-3m^2}{(2n-1)(2n+3)}\left[-\sigma\left(\frac{1}{2}+\delta\right) + \frac{2\pi b_P c_P J}{3\alpha}\left(Y_{10}^{*2}(\vartheta_P, \varphi_P) + Y_{11}^*(\vartheta_P, \varphi_P) Y_{1-1}^*(\vartheta_P, \varphi_P)\right)\right],$$

$$x_{n,m}^{\pm} = \sqrt{(1+n\pm m)(n\mp m)}\left\{(\gamma_1 \mp i\gamma_2)\frac{i\sigma h}{2\alpha}\right.$$
$$\left.\mp i\sqrt{\frac{\pi}{6}} b_P J Y_{1\pm 1}^*(\vartheta_P, \varphi_P) + \frac{\sqrt{2}\pi b_P c_P J(1\pm 2m)}{\alpha(2n-1)(2n+3)} Y_{10}^*(\vartheta_P, \varphi_P) Y_{1\pm 1}^*(\vartheta_P, \varphi_P)\right\},$$

$$x_{n,m}^{\pm\pm} = -\frac{\sqrt{(n\pm m+1)(n\pm m+2)(n\mp m-1)(n\mp m)}}{(2n-1)(2n+3)}\left(\frac{3\sigma}{4} + \frac{\pi b_P c_P J}{\alpha} Y_{1\pm 1}^{*2}(\vartheta_P, \varphi_P)\right),$$

$$y_{n,m} = \sqrt{\frac{(n+1)^2 - m^2}{(2n+1)(2n+3)}}\left\{-i\frac{m\sigma}{\alpha}\left(\frac{1}{2}+\delta\right) - n\sigma h\gamma_3 + \frac{nb_P J}{\alpha}\sqrt{\frac{\pi}{3}} Y_{10}^*(\vartheta_P, \varphi_P)\right.$$
$$\left.- i\frac{2\pi m b_P c_P J}{3}\left(Y_{10}^{*2}(\vartheta_P, \varphi_P) + Y_{11}^*(\vartheta_P, \varphi_P) Y_{1-1}^*(\vartheta_P, \varphi_P)\right)\right\},$$

$$y_{n,m}^{\pm} = \sqrt{\frac{(1+n\pm m)(2+n\pm m)}{(1+2n)(3+2n)}}\left\{\frac{nb_P J}{\alpha}\sqrt{\frac{\pi}{6}} Y_{1\pm 1}^*(\vartheta_P, \varphi_P)\right.$$
$$\left.\pm \frac{n}{2}\sigma h(\gamma_1 \mp i\gamma_2) \pm i\frac{\pi\sqrt{2} b_P c_P J(n\mp 2m)}{3} Y_{10}^*(\vartheta_P, \varphi_P) Y_{1\pm 1}^*(\vartheta_P, \varphi_P)\right\},$$

$$y_{n,m}^{\pm\pm} = \mp i\left(\frac{\sigma}{4\alpha} - \frac{\pi b_P c_P J}{3} Y_{1\pm 1}^{*2}(\vartheta_P, \varphi_P)\right)\sqrt{\frac{(1+n\pm m)(2+n\pm m)(3+n\pm m)(n\mp m)}{(1+2n)(3+2n)}},$$

$$w_{n,m} = \sqrt{\frac{n^2 - m^2}{4n^2 - 1}}\left\{-\frac{im\sigma}{\alpha}\left(\frac{1}{2}+\delta\right) + (n+1)\sigma h\gamma_3\right.$$
$$\left.- \frac{b_P J(n+1)}{\alpha}\sqrt{\frac{\pi}{3}} Y_{10}^*(\vartheta_P, \varphi_P) - i\frac{2m\pi b_P c_P J}{3}\left(Y_{10}^{*2}(\vartheta_P, \varphi_P) + Y_{11}^*(\vartheta_P, \varphi_P) Y_{1-1}^*(\vartheta_P, \varphi_P)\right)\right\},$$

$$w_{n,m}^{\pm} = \sqrt{\frac{(n\mp m)(n\mp m-1)}{4n^2-1}}\left\{\frac{b_P J(n+1)}{\alpha}\sqrt{\frac{\pi}{6}} Y_{1\pm 1}^*(\vartheta_P, \varphi_P)\right.$$
$$\left.\pm \frac{n+1}{2}\sigma h(\gamma_1 \mp i\gamma_2) \pm i\frac{\sqrt{2}\pi b_P c_P J(n+1\pm 2m)}{3} Y_{10}^*(\vartheta_P, \varphi_P) Y_{1\pm 1}^*(\vartheta_P, \varphi_P)\right\},$$

$$w_{n,m}^{\pm\pm} = \pm i\left(\frac{\sigma}{4\alpha} - \frac{\pi b_P c_P J}{3} Y_{1\pm1}^{*2}(\vartheta_P,\varphi_P)\right)\sqrt{\frac{(n\mp m-2)(n\mp m-1)(1+n\pm m)(n\mp m)}{4n^2-1}},$$

$$z_{n,m} = \frac{n}{2n+3}\sqrt{\frac{\left[(n+1)^2-m^2\right]\left[(n+2)^2-m^2\right]}{(2n+1)(2n+5)}}$$

$$\times\left\{\sigma\left(\frac{1}{2}+\delta\right) - \frac{2\pi b_P c_P J}{3\alpha}\left(Y_{10}^{*2}(\vartheta_P,\varphi_P) + Y_{11}^{*}(\vartheta_P,\varphi_P)Y_{1-1}^{*}(\vartheta_P,\varphi_P)\right)\right\},$$

$$z_{n,m}^{\pm} = -\frac{2\sqrt{2}\pi b_P c_P J}{3\alpha}Y_{10}^{*}(\vartheta_P,\varphi_P)Y_{1\pm1}^{*}(\vartheta_P,\varphi_P)\frac{n}{2n+3}\sqrt{\frac{\left[(n+1)^2-m^2\right](n\pm m+2)(n\pm m+3)}{(2n+1)(2n+5)}},$$

$$z_{n,m}^{\pm\pm} = -\left(\frac{\sigma}{4} + \frac{\pi b_P c_P J}{3\alpha}Y_{1\pm1}^{*2}(\vartheta_P,\varphi_P)\right)\frac{n}{2n+3}\sqrt{\frac{(n+1\pm m)(2+n\pm m)(3+n\pm m)(4+n\pm m)}{(2n+1)(2n+5)}},$$

$$v_{n,m} = \frac{n+1}{2n-1}\sqrt{\frac{\left[(n-1)^2-m^2\right](n^2-m^2)}{(2n+1)(2n-3)}}\left\{-\sigma\left(\frac{1}{2}+\delta\right)\right.$$

$$\left.+\frac{2\pi b_P c_P J}{3\alpha}\left(Y_{10}^{*2}(\vartheta_P,\varphi_P) + Y_{11}^{*}(\vartheta_P,\varphi_P)Y_{1-1}^{*}(\vartheta_P,\varphi_P)\right)\right\},$$

$$v_{n,m}^{\pm} = -\frac{2\sqrt{2}\pi b_P c_P J}{3\alpha}Y_{10}^{*}(\vartheta_P,\varphi_P)Y_{1\pm1}^{*}(\vartheta_P,\varphi_P)\frac{n+1}{2n-1}\sqrt{\frac{(n\mp m-2)(n\mp m-1)(n^2-m^2)}{(2n+1)(2n-3)}},$$

$$v_{n,m}^{\pm\pm} = \left(\frac{\sigma}{4} + \frac{\pi b_P c_P J}{3\alpha}Y_{1\pm1}^{*2}(\vartheta_P,\varphi_P)\right)\frac{n+1}{2n-1}\sqrt{\frac{(n\mp m-3)(n\mp m-2)(n\mp m-1)(n\mp m)}{(2n+1)(2n-3)}}.$$

In order to rewrite Eq. (42) in the form of Eq. (20) explicitly, we define $\mathbf{C}_n(t)$ as the column vectors arranged in an appropriate way from $c_{n,m}(t)$, viz.,

$$\mathbf{C}_0(t) = \mathbf{0}, \quad \mathbf{C}_n(t) = \begin{pmatrix} c_{2n,-2n}(t) \\ c_{2n,-2n+1}(t) \\ \vdots \\ c_{2n,2n}(t) \\ c_{2n-1,-2n+1}(t) \\ c_{2n-1,-2n+2}(t) \\ \vdots \\ c_{2n-1,2n-1}(t) \end{pmatrix}, \quad (n \geq 1), \tag{43}$$

while the matrices $\mathbf{Q}_n, \mathbf{Q}_n^+, \mathbf{Q}_n^-$ are defined as

$$\mathbf{Q}_n = \begin{pmatrix} \mathbf{X}_{2n} & \mathbf{W}_{2n} \\ \mathbf{Y}_{2n-1} & \mathbf{X}_{2n-1} \end{pmatrix}, \quad \mathbf{Q}_n^+ = \begin{pmatrix} \mathbf{Z}_{2n} & \mathbf{Y}_{2n} \\ \mathbf{0} & \mathbf{Z}_{2n-1} \end{pmatrix}, \quad \mathbf{Q}_n^- = \begin{pmatrix} \mathbf{V}_{2n} & \mathbf{0} \\ \mathbf{W}_{2n-1} & \mathbf{V}_{2n-1} \end{pmatrix}. \tag{44}$$

In turn, the matrices $\mathbf{Q}_n, \mathbf{Q}_n^+, \mathbf{Q}_n^-$ themselves consist of five submatrices $\mathbf{V}_l$, $\mathbf{W}_l$, $\mathbf{X}_l$, $\mathbf{Y}_l$, and $\mathbf{Z}_l$ of dimensions $(2l+1)\times(2l-3)$, $(2l+1)\times(2l-1)$, $(2l+1)\times(2l+1)$, $(2l+1)\times(2l+3)$, and $(2l+1) \times (2l+5)$, respectively. The elements of these five-diagonal submatrices, which are formed from the coefficients occurring in Eq. (42), are given by

$$\left(\mathbf{V}_l\right)_{n,m} = \delta_{n-4,m} v^{--}_{l,-l+m+3} + \delta_{n-3,m} v^{-}_{l,-l+m+2} + \delta_{n-2,m} v_{l,-l+m+1} + \delta_{n-1,m} v^{+}_{l,-l+m} + \delta_{n,m} v^{++}_{l,-l+m-1},$$

$$\left(\mathbf{W}_l\right)_{n,m} = \delta_{n-3,m} w^{--}_{l,-l+m+2} + \delta_{n-2,m} w^{-}_{l,-l+m+1} + \delta_{n-1,m} w_{l,-l+m} + \delta_{n,m} w^{+}_{l,-l+m-1} + \delta_{n+1,m} w^{++}_{l,-l+m-2},$$

$$\left(\mathbf{X}_l\right)_{n,m} = \delta_{n-2,m} x^{--}_{l,-l+m+1} + \delta_{n-1,m} x^{-}_{l,-l+m} + \delta_{n,m} x_{l,-l+m-1} + \delta_{n+1,m} x^{+}_{l,-l+m-2} + \delta_{n+2,m} x^{++}_{l,-l+m-3},$$

$$\left(\mathbf{Y}_l\right)_{n,m} = \delta_{n-1,m} y^{--}_{l,-l+m} + \delta_{n,m} y^{-}_{l,-l+m-1} + \delta_{n+1,m} y_{l,-l+m-2} + \delta_{n+2,m} y^{+}_{l,-l+m-3} + \delta_{n+3,m} y^{++}_{l,-l+m-4},$$

$$\left(\mathbf{Z}_l\right)_{n,m} = \delta_{n,m} z^{--}_{l,-l+m-1} + \delta_{n+1,m} z^{-}_{l,-l+m-2} + \delta_{n+2,m} z_{l,-l+m-3} + \delta_{n+3,m} z^{+}_{l,-l+m-4} + \delta_{n+4,m} z^{++}_{l,-l+m-5}.$$